\documentclass{aa}

\usepackage{graphicx}

\begin{document}

\newcommand{\be}{\begin{equation}}
\newcommand{\ee}{\end{equation}}

\title{An efficient shock-capturing central-type scheme 
  for multidimensional relativistic flows}
\subtitle{I. Hydrodynamics}

\author{L. Del Zanna    
\and    N. Bucciantini} 

\offprints{L. Del Zanna; \\
\email{ldz@arcetri.astro.it}}

\institute{Dipartimento di Astronomia e Scienza dello Spazio,
             Universit\`a degli Studi di Firenze, \\
             Largo E. Fermi 2, I-50125 Firenze, Italy}

\date{Received 7 March 2002; accepted ...}

\abstract{
Multidimensional shock-capturing numerical schemes for special 
relativistic hydrodynamics (RHD) are computationally more expensive 
than their correspondent Euler versions, due to the nonlinear
relations between conservative and primitive variables and to the consequent
complexity of the Jacobian matrices (needed for the spectral decomposition 
in most of the approximate Riemann solvers of common use). 
Here an efficient and easy-to-implement three-dimensional (3-D) 
shock-capturing scheme for ideal RHD is presented.
Based on the algorithms developed by P. Londrillo and L. Del~Zanna 
({\em Astrophys. J.} 530, 508-524, 2000) for the non-relativistic 
magnetohydrodynamic (MHD) case, and having in mind its relativistic MHD
extension (to appear in a forthcoming paper), the scheme uses high order
(third) Convex Essentially Non-Oscillatory (CENO) finite difference 
interpolation routines and central-type averaged Riemann solvers, which
do not make use of time-consuming characteristic decomposition.
The scheme is very efficient and robust, and it gives results comparable
to those obtained with more sophisticated algorithms, even in 
ultrarelativistic multidimensional test problems.
}

\authorrunning{L.~Del Zanna and N. Bucciantini}
\titlerunning{A shock-capturing scheme for relativistic flows
               -- I. Hydrodynamics}

\maketitle


\section{Introduction}

Relativistic flows and shocks play an essential role in modern high energy
astrophysics, both for the interpretation of various observed features and
for the description of the physical processes they give rise to, as,
for example, the acceleration of highly energetic particles.
Among the various astrophysical objects in which relativistic flows 
have been invoked to explain the observed properties, 
the best studied are probably: 
\begin{enumerate}
\item
Active galactic nuclei (AGNs), for which the presence of relativistic bulk
motions (up to Lorentz factors of order 10) was soon suggested (Rees 1967).
Associated to AGNs are often highly collimated relativistic jets
(Begelman et al. 1984; and, for a review, Ferrari 1998), seen as apparent
super-luminal lobes in some radio-loud sources and whose shock fronts
are among the most promising candidates for high-energy particle acceleration. 
Mildly relativistic jets appear to be also associated to galactic X-ray 
compact sources (generally called {\em microquasars}, see Mirabel 
\& Rodriguez 1999, for a review).
\item
Gamma-ray bursts (GRBs), supposed to be originated during the collapse 
of the iron core of massive stars and the subsequent {\em fireball} 
explosion (Piran 1999; Kobayashi et al. 1999), which give rise to an 
expanding blast wave of pairs and hadrons with typical Lorentz factors of 
$10^2-10^3$ (Meszaros \& Rees 1992), whose kinetic energy is then believed 
to be converted into gamma rays via cyclotron radiation and/or
inverse Compton scattering.
\item
Pulsar wind nebulae, assumed to be bubbles of relativistic particles and
magnetic fields emitted by a pulsar as a relativistic wind with Lorentz factor
ranging from $10^4$ to $10^7$, depending on the model of pair production in the
pulsar magnetosphere (Michel \& Li 1999). 
The wind region may be confined by a termination shock, generated by
the interaction with outer supernova matter, as is the case for the
synchrotron emitting nebulae called plerions (Rees \& Gunn 1974;
Kennel \& Coroniti 1984); or, if the pulsar is moving with supersonic speed
through the interstellar medium, by the resulting ram pressure, 
in this case giving rise to a {\em bow-shock} pulsar wind nebula that
may be detected in H$\alpha$ (Bucciantini \& Bandiera 2001).
\end{enumerate}
As we can see from these few but important examples, there is a strong 
interest in the astrophysical community to the development of computational 
codes for the numerical modeling and simulation of relativistic flows.

Over the last decade, high resolution shock-capturing methods
of Godunov type, successfully applied in classical fluid dynamics,
have started to be employed for the case of relativistic hydrodynamics as well
(Marquina et al. 1992; Schneider et al. 1993; Balsara 1994; Duncan \&
Hughes 1994; Eulderink \& Mellema 1994;  Font et al. 1994; 
Dolezal \& Wong 1995; Falle \& Komissarov 1996; Donat et al. 1998;
Aloy et al. 1999).
These schemes are characterized by the following main features: 
a conservative form of the discretized equations, in order to capture 
weak solutions and satisfy jump relations; a reconstruction phase, 
to recover variables at inter-cell locations where fluxes have to be 
computed; an upwind phase, in which an exact or approximate solution 
to the local Riemann problem is found.
The simulated flows achieve high accuracy in smooth regions and, at the same
time, shock profiles are stable and sharply defined. 
This is the reason for the great success of {\em upwind} differencing 
over central (or spectral) differencing. In the latter case,
artificial viscosity must be introduced in order to damp the
spurious oscillations that always form near discontinuities (Gibbs 
phenomena), thus leading to artificial heating and unwanted damping 
of physical waves.

When discontinuous solutions are of main interest, second order
(both in time and space) total variation diminishing (TVD) schemes 
coupled with an accurate Riemann solver are probably the best choice, 
in terms of sharp resolution of discontinuities and overall stability.
However, in the general case smooth wave-like or even turbulent fields
appear together with discontinuous solutions, thus second order schemes
are no longer able to resolve both features with enough precision. 
By relaxing the stringent TVD condition, a class of essentially non-oscillatory 
(ENO) higher-order schemes were first proposed by Harten et al. (1987), 
later modified in a more efficient implementation by Shu \& Osher (1989).
The ENO philosophy is to use {\em adaptive} stencils to reconstruct
variables and fluxes at cell interfaces. 
Thus, in smooth regions symmetric stencils will be used,
whereas near discontinuities the stencil will shift to the
left or to the right, selecting the smoother part of the flow and
thus achieving the same high resolution (typically from third to fifth order
in the case of weighted ENO schemes, see Jiang \& Shu 1996) everywhere,
without the need of expensive adaptive mesh refinement (AMR) techniques.
The price to pay is the presence, near shocks, of small oscillations of
the order of the truncation error (Gibbs oscillations given by centered
stencils that cross a discontinuity would be much higher, since they are
proportional to the jump itself, independently on the resolution). 

Like for most TVD Godunov-type schemes, also for ENO algorithms the
local spectral decomposition in the building of numerical fluxes
is commonly adopted. It was strongly advocated already in the
original paper, where a version of the ENO scheme with the (much) 
simpler component-wise reconstructions was found to fail, giving
significant oscillations, in some Riemann problems. However, due
to the high resolution of ENO methods, characteristic decomposition
on every point of the interpolation stencil becomes prohibitive
when moving to three-dimensional simulations, especially for relativistic
flows where Jacobian matrices are more complex than in the Eulerian case.  
The same kind of problem has to be faced in magnetohydrodynamics 
(MHD, see Londrillo \& Del Zanna 2000, from now on LD), 
because of the increasing number of variables, equations 
and eigenmodes. The worst possible case is obviously that of relativistic
MHD, for which only second order 1-D (Balsara 2001) and 2-D 
(Komissarov 1999) Godunov schemes have been presented so far.

Following the scheme proposed in LD, in the present series of two papers
a shock-capturing scheme that avoids the expensive characteristic
decomposition, still retaining non-oscillatory properties, will be suggested, 
first for ideal relativistic hydrodynamics (RHD, in this paper) 
and then for ideal relativistic magnetohydrodynamics (RMHD, in the 
next paper of the series). Our method is based on the so-called
{\em central} schemes, that extend the first-order local Lax-Friedrichs
(LLF) flux splitting and assume an average over the Riemann fan at every
cell interface (see Nessyahu \& Tadmor 1990, for the original paper
and Kurganov et al. 2001, for the latest developments in this field).
The basic properties of our method can be found in Liu \& Osher (1998),
where a third order multidimensional central scheme was presented, with
the following main features:
\begin{enumerate}
\item Point values rather than cell averages are used (that is to say 
finite differences instead of finite volumes, see Sect.~3 for details),
thus removing the need of staggered grids (employed in previous central
schemes) and making the extension to multiple dimensions a trivial task.
\item  The semi-discrete form of the equations is solved, so that
time integration can be achieved with any solver for ordinary differential 
equations (ODEs), for example TVD Runge-Kutta methods.
\item No characteristic decomposition and Riemann solvers are required:
fluxes are reconstructed and derived component-wise, thus achieving
a great simplicity in the programming and, above all, efficiency.
The only spectral pieces of information needed are the local highest 
characteristic speeds (related to the Courant coefficient).
\item A new third order reconstruction algorithm is introduced and tested,
called Convex-ENO (CENO), which has the fundamental property to stay as
close as possible to the lower order TVD limited linear reconstruction 
(by using the {\em minmod} limiter in this case), 
thus reducing even to first order where needed, while retaining
high accuracy in smooth region. This was recognized to be the key point 
of the surprising success of central schemes (Tadmor, cited in their
acknowledgements), confirmed for the MHD case by LD.
\end{enumerate}
Here, like in LD, we take advantage of all these positive features and 
we modify the original scheme by splitting the CENO reconstruction into 
two separate routines, so that reconstruction can be applied to primitive
variables rather than to fluxes (the resulting scheme appears to be more
robust, especially in the relativistic case). Moreover, different limiters
(e.g. the {\em monotonized centered}) and solvers (the upwind HLL,
from Harten, Lax, and van Leer) are introduced and tested in our framework.

We will show that our scheme, by providing an overall high resolution,
is able to compensate for the larger smearing of discontinuities, especially
contact discontinuities (due to the use of solvers based on just one or 
two characteristic speeds) and the results are comparable with those
obtained by more sophisticated (but much more computationally expensive,
especially in 3-D) relativistic codes. 
Concluding, notice that the proposed scheme can be applied unchanged to
Euler HD too, where only the definition of conservative variables, fluxes
and characteristic speeds are different.

\section{Ideal RHD equations}

The covariant equations in special relativistic hydrodynamics (RHD) 
are (e.g. Landau \& Lifshitz 1959):
\be
\partial_\alpha (\rho u^\alpha ) = 0,
\label{mass}
\ee
\be
\partial_\alpha (w u^\alpha \! u^\beta + pg^{\alpha\beta} ) = 0,
\label{momentum}
\ee
where velocities are normalized against the speed of light ($c=1$) and
Greek letters indicate four-vectors, while Latin indexes will be devoted
to spatial 3-D vectors. The metric tensor defines space-time properties 
and here a Minkowski flat space with $g^{\alpha\beta}=\mathrm{diag}
\{-1,1,1,1\}$ will be assumed throughout for simplicity, with coordinates
$x^\alpha=(t,x^j)$. The physical quantities involved in the
conservation laws are the rest mass density $\rho$, the kinetic pressure
$p$, the relativistic enthalpy $w=e+p$ ($e$ is the energy per unit volume,
including rest mass energy) and the four-velocity $u^\alpha=(\gamma,
\gamma v^j)$, where $\gamma=(1-v^2)^{-1/2}$ is the Lorentz factor.
The system has to be closed with an equation of state $p=p(\rho,e)$, and
here the relation for an ideal gas
\be
p=(\Gamma-1)(e-\rho)\Rightarrow e=\rho+p/(\Gamma-1)
\label{gamma}
\ee
will be considered, where the adiabatic index should be taken as $\Gamma=5/3$ 
for the mildly relativistic case and as $\Gamma=4/3$ for the ultrarelativistic
case $e\gg\rho$.

Godunov-type shock-capturing numerical methods developed
for classical Euler equations can actually be applied
to any multidimensional system of hyperbolic conservation laws of the form 
\be
\frac{\partial \vec{u}}{\partial t} + 
\sum_{i=1}^{d} \frac{\partial \vec{f}^i(\vec{u})}{\partial x^i} = 0,
\label{cons_law}
\ee
where $\vec{u}$ is the vector of conserved variables and $\vec{f}^i$
are their corresponding fluxes, along each direction ($d$ is the number
of spatial dimensions). 
Equations (\ref{mass}) and (\ref{momentum}) are automatically cast in
this form by just defining 
\be
\vec{u}(\vec{v})=[\rho\gamma,w\gamma^2 v^j,w\gamma^2-p]^T,
\label{cons}
\ee
\be
\vec{f}^i(\vec{v})=[\rho\gamma v^i,w\gamma^2 v^i\! v^j+p\delta^{ij},
w\gamma^2 v^i]^T,
\label{flux}
\ee
where $\vec{v}=[\rho,v^j,p]^T$ are called primitive variables, and
therefore the numerical techniques largely used for Euler HD equations
can be applied to RHD too.
The hyperbolic nature of Eq.~(\ref{cons_law}) is guaranteed provided the local
sound velocity $c_s$ is sub-luminal, that is for causal equations of state
(Anile 1989). For $\Gamma$-law gases where Eq.~(\ref{gamma}) holds,
the relativistic sound speed is given by
\be
c_s^2=\left(\frac{\partial p}{\partial e}\right)_s\!\!\!=
\left(\frac{\partial e}{\partial p}\right)_s^{-1}\!\!\!=
a^2\left(1+\frac{a^2}{\Gamma-1}\right)^{-1}\!\!\!\equiv\Gamma\,p/w,
\label{sound}
\ee
which is obviously always less than unity, where $s\sim p\rho^{-\Gamma}$
is the specific entropy and $a^2=\Gamma p\rho^{-1}$ defines the classical 
sound speed.

In any numerical time advancing routine, primitive variables $\vec{v}$ have 
to be derived from the conservative ones at least once per time step, 
and if this is trivial for Euler equations it is not so in the RHD case, 
in which a numerical nonlinear root-finding technique must be employed.
If we define $W=w\gamma^2$ and
\be
\vec{u}=[D,Q^j,E]^T,
\ee
where $D=\rho\gamma$, $Q^j=Wv^j$, and $E=W-p$, the system to be inverted
can be cast in the single equation for $\gamma$:
\be
W(\gamma)^2(1-\gamma^{-2})-Q^2=0,
\ee
with $w=\rho+\Gamma_1p\Rightarrow W=D\gamma+\Gamma_1\gamma^2p$, here
$\Gamma_1=\Gamma/(\Gamma-1)$, and $p=W-E$, so that
\be
W(\gamma)=\frac{E\Gamma_1\gamma^2-D\gamma}{\Gamma_1\gamma^2-1}.
\label{enthalpy}
\ee
Once the Lorentz factor is found numerically with the requested 
accuracy, the primitive variables are easily recovered thanks to
the relations above.

\section{A novel ENO-based central scheme}

In this section ENO methods and their implementation for component-wise 
central schemes, together with our specific modifications, will be presented. 
For a general introduction and review of ENO methods for hyperbolic 
conservation laws see Shu (1997).

Consider a numerical discretization of Eq.~(\ref{cons_law}),
in the one-dimensional case $d=1$ to begin with. Given an interval
$[a,b]$, $N$ numerical cells $I_i=[x_{i-1/2},x_{i+1/2}]$ of equal
length $\Delta x=(b-a)/N$ can be defined, with cell centers (grid
nodes) given by
\be
x_i=a+(i-1/2)\,\Delta x;~~i=1,\ldots,N.
\ee
To an order $r$ of spatial accuracy, the numerical value of any 
quantity $v(x)$ (at a given time) will be denoted as $v_i=v(x_i)
+O({(\Delta x})^r)$ at grid points (the so-called {\em point values})
or $v_{i\pm 1/2}=v(x_{i\pm 1/2})+O(({\Delta x})^r)$ at cell boundaries,
where the order of accuracy refers only to smooth regions, in which the
larger stencil can be used for interpolation.
Moreover, cell averages are defined as
\be
\bar{v}_i=\frac{1}{\Delta x}\int_{x_{i-1/2}}^{x_{i+1/2}}v(x){\rm d}x,
\label{cell_av}
\ee
and only for schemes up to second order, $r\leq 2$, they do coincide with 
point values $v_i$.

Most of the shock-capturing schemes evolve the cell averaged conserved 
quantities $\bar{\vec{u}}_i$ in time, as obtained by integrating
Eq.~(\ref{cons_law}) over the cell $I_i$. 
However, in the multidimensional case, the resulting numerical fluxes 
$\hat{\vec{f}}_i$, that discretize the physical ones along one direction, 
have to be averaged along the transverse directions. This implies a truly
multidimensional numerical interpolation for high order schemes, which
is usually complex and computationally expensive. This approach is
generally referred to as the {\em finite volume} approximation.
Like in LD and in many other ENO schemes, from the works by Shu \& Osher 
(1988; 1989) onwards, we will adopt here {\em finite difference}
approximations, based on point values. In the semi-discrete formalism 
(that is retaining continuous time dependency in the spatially discretized 
quantities), Eq.~(\ref{cons_law}) becomes ($d=1$):
\be
\frac{{\rm d}\vec{u}_i}{{\rm d}t}=
-\frac{\hat{\vec{f}}_{i+1/2}-\hat{\vec{f}}_{i-1/2}}{\Delta x},
\label{semi}
\ee
where $\hat{\vec{f}}_{i\pm 1/2}$ are high-order approximations to the 
{\em primitives} of physical fluxes, that is to say that cell averages of the 
$\hat{\vec{f}}(x)$ function must coincide with point values $\vec{f}_i$ 
of the flux function $\vec{f}(x)$, to the given accuracy 
($\hat{\vec{f}}_{i\pm 1/2}\equiv\vec{f}_{i\pm 1/2}$ up to second order).
The extension to the multidimensional case is now straightforward, 
since interpolations from cell centers to cell boundaries are made
separately, dimension by dimension, and the other derivatives are just
subtracted from the right hand side of Eq.~(\ref{semi}) exactly in
the same fashion as for the discretized $x$ derivative.

As specified by Shu \& Osher (1988), Eq.~(\ref{semi}) must be integrated
in time by using proper multi-level Runge-Kutta methods corresponding to the
high order of spatial accuracy, so here we will always employ the optimal
third order TVD algorithm:
\begin{eqnarray}
\vec{u}^{(1)} & = & \vec{u}^n+\Delta t\,\mathcal{L}[\vec{u}^n] \nonumber \\
\vec{u}^{(2)} & = & \frac{3}{4}\vec{u}^n+\frac{1}{4}\vec{u}^{(1)}+
\frac{1}{4}\Delta t\,\mathcal{L}[\vec{u}^{(1)}] \\
\vec{u}^{n+1} & = & \frac{1}{3}\vec{u}^n+\frac{2}{3}\vec{u}^{(2)}+
\frac{2}{3}\Delta t\,\mathcal{L}[\vec{u}^{(2)}] \nonumber
\end{eqnarray}
where the superscript $n$ indicates the time stepping discretization,
$\vec{u}^{(1)}$ and $\vec{u}^{(2)}$ 
refer to intermediate integration stages, and $\mathcal{L}[\vec{u}]$ 
is simply the right hand side of Eq.~(\ref{semi}). 
The explicit time advancing scheme
above is stable under the CFL (Courant-Friedrichs-Lewy) condition
$c<1$, where the Courant coefficient appears in the definition of the
maximum time step allowed:
\be
\Delta t = \frac{c}{\mathrm{max}_i(\alpha^i/\Delta x^i)},
\label{cfl}
\ee
with $\alpha^i$ being the largest speed of propagation of characteristic waves
in the direction $i$. In the relativistic case, there is clearly a lower
limit for the time step, given in Eq.~(\ref{cfl}) by simply taking 
$\alpha^i=1$ along all directions.

In order to complete the description of the scheme, the interpolation
techniques and the approximate Riemann solver that defines the numerical
inter-cell fluxes $\hat{\vec{f}}_{i+1/2}$ must be given.
The following steps are taken, for every Runge-Kutta sub-cycle and
for every direction in the multidimensional case:

\begin{enumerate}

\item
Primitive variables are recovered from the conservative ones, according
to the recipe already given in the previous section (this step is actually
taken just once for each sub-cycle):
\be
\{\vec{u}_i\}\longrightarrow \{\vec{v}_i\};~~i=1,\ldots,N.
\label{step1}
\ee

\item
Primitive variables are reconstructed at cell interfaces to give two
states, called left and right inter-cell states:
\be
\{\vec{v}_i\}\longrightarrow \{\vec{v}^L_{i+1/2}\},\,\{\vec{v}^R_{i+1/2}\};
~~i=0,\ldots,N.
\label{step2}
\ee
The two reconstructions are based on polynomial interpolation over different 
sets of stencils, one centered on $x_i$ to approximate the left state
$\vec{v}^L_{i+1/2}$ and the other centered on $x_{i+1}$ to approximate 
the right state $\vec{v}^R_{i+1/2}$. 
The interpolation is performed separately on each variable $v(x)$ according
to a {\em Rec} routine based on the CENO (Convex ENO, Liu \& Osher 1998)
technique, as described in Appendix~A. Contrary to common ENO schemes,
our method uses point values $v_i$ to yield point value reconstructed 
quantities $v_{i+1/2}$, instead of starting from cell averages $\bar{v}_i$.
The order of the reconstruction is $r=3$ in smooth regions, but it
reduces to linear reconstruction or even to first order (by using 
{\em minmod}-type limiters) near discontinuities. In this latter case
$v^L_{i+1/2}=v_i$ and $v^R_{i+1/2}=v_{i+1}$, for each discontinuous field.

\item
At each inter-cell point $x_{i+1/2}$ the local Riemann problem must be
solved, in some approximate way. The most accurate solution would be given
by an exact solver that computes the evolved state from $\vec{v}^L$
and $\vec{v}^R$ and then the correspondent flux, to be finally identified
with $\vec{f}_{i+1/2}$. Another possibility is to define an
average state $\tilde{\vec{v}}_{i+1/2}$ and to decompose fluxes according
to a linearized problem with the Jacobian $\partial\vec{f}/\partial\vec{u}$
calculated there (Roe matrix approach). Our choice is to avoid spectral
decomposition, which is computationally expensive, and to take an average
over the local Riemann fan as in central-type schemes, already mentioned
in the introduction. 
A simple two-speeds Riemann solver is the HLL one, first used in central
schemes by Kurganov et al. (2001),
that retains an upwind nature in the sense that coincides
with $\vec{f}^L$ or $\vec{f}^R$ if the Riemann fan does not cross the
inter-cell itself (all the eigenvalues have the same sign).
This may be written in the form
\be
\vec{f}^{\rm HLL}=\frac{\alpha^+\vec{f}^L+\alpha^-\vec{f}^R
-\alpha^+\alpha^-(\vec{u}^R-\vec{u}^L)}{\alpha^++\alpha^-},
\label{hll}
\ee
where all quantities are calculated from the reconstructed values of
Eq.~(\ref{step2}) by using Eqs.~(\ref{cons}) and (\ref{flux}). Here
the $\alpha^\pm$ coefficients take into account the highest speeds at the
two sides of the Riemann fan, which can be estimated from the maximum
and minimum eigenvalue $\lambda^\pm$ of the Jacobians at the left and right
states:
\be
\alpha^\pm=\mathrm{max}\{0,\pm\lambda^\pm(\vec{v}^L),
\pm\lambda^\pm(\vec{v}^R)\}.
\label{alpha_pm}
\ee
For relativistic flows, the required eigenvalues are (after splitting the
velocity along and perpendicular to the direction of spatial derivation,
according to the relativistic rule for addition of velocity vectors):
\be
\lambda^\pm=\frac{v_\parallel (1-c_s^2)\pm c_s\sqrt{(1-v^2)
(1-v_\parallel^2-v^2_\perp c_s^2)}}{1-v^2c_s^2},
\label{lambda}
\ee
that reduce simply to $\lambda^\pm=(v\pm c_s)/(1\pm v c_s)$ in the 
one-dimensional case. Note that the maximum and minimum eigenvalues
at the two reconstructed states are the only spectral pieces of information
required. In this way shocks are handled correctly, whereas contact
discontinuities and shear waves, corresponding to intermediate eigenvalues
$\lambda=v_\parallel$, can appear somehow smeared, when compared with 
the results from proper Riemann solvers. 

The simplest, smoothest, but also most dissipative numerical flux 
is the (local) Lax-Friedrichs one, given by
\be
\vec{f}^{\rm LLF}=\frac{1}{2}
[\vec{f}^L+\vec{f}^R-\alpha (\vec{u}^R-\vec{u}^L)],
\label{llf}
\ee
in which $\alpha=\mathrm{max}\{\alpha^+,\alpha^-\}$ and therefore
the averaging region is always symmetric with respect to $x_{i+1/2}$
(the LLF numerical flux is the prototype of central schemes).
Note that the maximum of the set of values $\{\alpha_{i+1/2}\}$ yields
the CFL-related $\alpha$ coefficient appearing in Eq.~(\ref{cfl}), and
actually they coincide for the so-called global LF numerical flux.
The calculation of eigenvalues may be avoided completely in a global
LF scheme that uses $\alpha=1$ everywhere, although this is the most 
smearing case. Note that Eq.~(\ref{hll}) reduces to Eq.~(\ref{llf}) when
$\alpha^+=\alpha^-=\alpha$, that is for symmetric Riemann fans in which
$v_\parallel=0\Rightarrow\lambda^+=-\lambda^-$.

\item
From the point values of numerical fluxes, the approximations of their
derivatives must be finally calculated (for lower than third order
schemes this step can be avoided):
\be
\{\vec{f}_{i+1/2}\}\longrightarrow \{\hat{\vec{f}}_{i+1/2}\};
~~i=0,\ldots,N.
\label{step4}
\ee
These are the numerical fluxes that actually enter Eq.~(\ref{semi}).
This last step usually does not appear in other ENO high-order schemes,
since the reconstruction is made directly on fluxes and steps 2 and 4
are taken simultaneously (the numerical flux is calculated before
reconstruction, using {\em flux splitting} methods). However, we have
noticed that for central-type schemes like the one presented here, which
avoid characteristic decomposition, the proposed approach is more robust
and less oscillatory. 
This final interpolation is again performed separately on each variable 
with another CENO-based routine ({\em Der}, see Appendix~A).

\end{enumerate}

\section{Numerical results}

In this section the proposed scheme is validated against typical tests
available in the current literature, separated here in one-dimensional
tests, essentially shock-tube Riemann problems, and a few 2-D and 3-D
experiments. All the simulations have been run on a single PC-Linux
(AMD Athlon 1GHz CPU, 512Mb RAM, Pacific-Sierra Vast/f90 compiler,
--O2 optimization and single precision calculations) in
order to demonstrate that our scheme is not particularly demanding in
terms of computational resources. In 3-D, with a grid of $100^3$ nodes,
typical speeds are of about 2 mins per Runge-Kutta iteration (three
internal sub-cycles for the third order scheme), so a simulation like
the spherical explosion takes a few hours of computational time.
However, the code is fully parallelized with 
{\em Message Passing Interface} (MPI) directives and has been
successfully tested on a variety of supercomputers, like CrayT3E, IBM SP3/4 
and Beowulf clusters.

In the following tests, we will always use the proposed third order CENO3 
as a base scheme, unless specified otherwise. The approximate Riemann
solvers employed are the HLL or the LLF, as described in the previous
section, whereas the slope limiters used in the linear part of the 
reconstruction (see Appendix~A) are the Monotonized Centered (MC) or the 
MinMod (MM). All the simulations presented here use the classical value
$\Gamma=5/3$ for the adiabatic index.

\subsection{One-dimensional tests}

Shock-capturing numerical schemes, as in the classical hydrodynamical case,
must be able to reproduce the discontinuous profiles involved in the
so-called {\em shock-tube} problems. Given a unitary numerical 1-D {\em pipe}
of $N$ grid points, two constant states $(\rho,v,p)$ are taken on the
left ($0\leq x \leq 0.5$) and on the right ($0.5< x \leq 1$) with respect
of a diaphragm, placed initially at $x=0.5$ and then removed.
Typical patterns seen in the subsequent evolution are shocks, contact
discontinuities (characterized by density jumps not accompanied by 
discontinuities in normal velocity and pressure) and rarefaction waves.
In the relativistic regime these features are qualitatively unchanged,
since the structure of the characteristics is the same, but density
jumps are not limited by any function of the adiabatic index and
rarefaction waves do not yield straight profiles, due to the nonlinear
Lorentz transformation formulae.

\begin{figure}
\resizebox{\hsize}{!}{\includegraphics{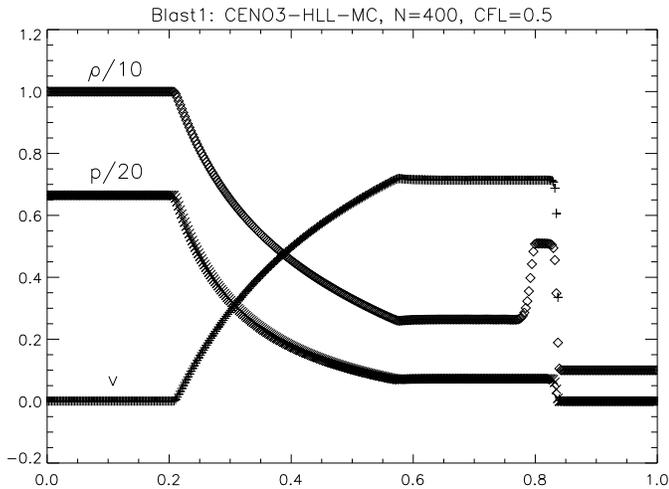}}
\caption{The relativistic blast wave problem 1 for time $t=0.4$.
The computed profiles of density (diamond), pressure (cross), and
velocity (plus) are shown against distance. The base third order CENO3 
scheme is employed with MC limiter and HLL solver. The computational grid 
is formed by $N=400$ equidistant cells and the Courant number is $c=0.5$.}
\label{fig1}
\end{figure}

\begin{figure}
\resizebox{\hsize}{!}{\includegraphics{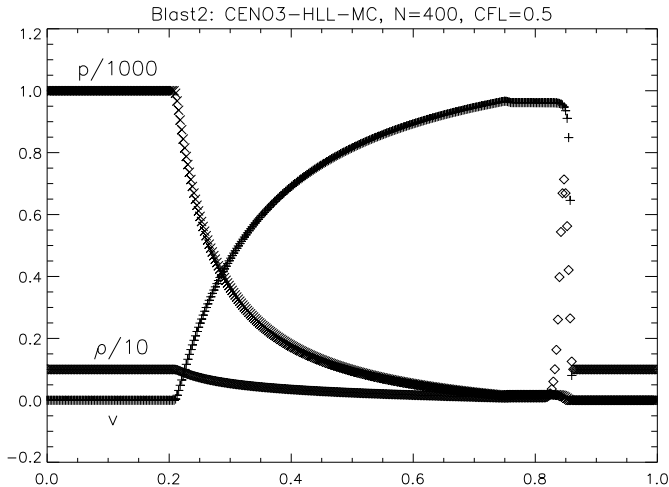}}
\caption{The relativistic blast wave problem 2 for time $t=0.35$.
The computed profiles of density (diamond), pressure (cross), and
velocity (plus) are shown against distance. The base third order CENO3 
scheme is employed with MC limiter and HLL solver. The computational grid 
is formed by $N=400$ equidistant cells and the Courant number is $c=0.5$.
The exact value for the density peak is around $10.5$, our numerical
result is around $7.3$, similar to what is found by other third order 
more sophisticated schemes.}
\label{fig2}
\end{figure}

\begin{figure}
\resizebox{\hsize}{!}{\includegraphics{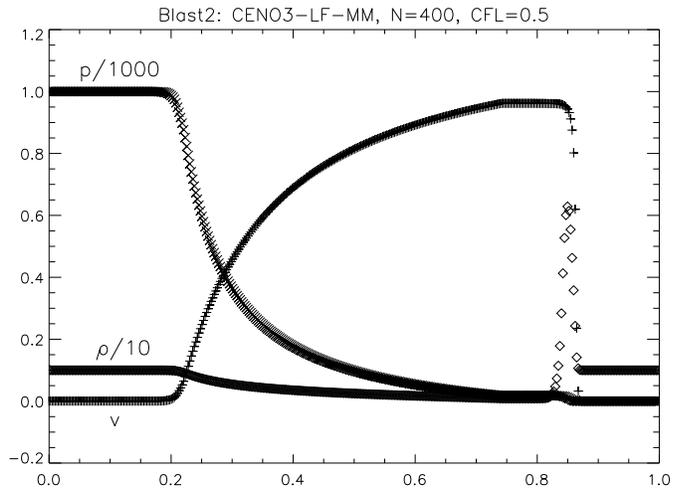}}
\caption{The same problem and settings as in Fig.~2, except for limiter
and solver. Here the most smearing case of MM limiter and global LF flux
splitting with $\alpha=1$ (highest possible value for relativistic flows)
is tested. Differences arise only at $x\simeq 0.2$ and $x\simeq 0.75$, 
where the changes of slope due to the rarefaction wave are less sharp, and 
at the density peak, which is lower ($\rho\simeq 6.5$) in this case.}
\label{fig3}
\end{figure}

First we present two relativistic {\em blast wave} explosion problems,
characterized by an initial static state with temperature and pressure
much higher in the region on the left:
$$
\mbox{Blast wave 1:}
\left\{ \begin{array}{ll} 
(\rho,v,p)^L = & (10,0,13.3), \\
(\rho,v,p)^R = & (1,0,10^{-6}),
\end{array}\right.
$$
$$
\mbox{Blast wave 2:}
\left\{ \begin{array}{ll} 
(\rho,v,p)^L = & (1,0,1000), \\
(\rho,v,p)^R = & (1,0,0.01),
\end{array}\right.
$$
as in the Donat et al. (1998) paper. The first test is only mildly
relativistic, while the second is more severe, with a shock speed corresponding
to $\gamma\simeq 6$. In Fig.~\ref{fig1} and  Fig.~\ref{fig2} the two
tests are simulated with $N=400$ and Courant number $c=0.5$, with the
base scheme CENO3-HLL-MC. Note the total absence of oscillations and
the accuracy in the definition of shocks and rarefaction waves.
The contact discontinuity is more smeared, due to the use of a solver
that does not take into account that intermediate wave and above
all to the reconstruction method, which is the same for all the quantities,
thus we cannot steepen (e.g. with the {\em superbee} limiter) the contact
discontinuity alone, as usually done in characteristics based schemes.
In particular, the height of the density peak in Fig.~\ref{fig3} provides
a measure of the numerical viscosity of the scheme: the exact value should
be around $10.5$, while here we find about $7.3$, slightly more than what
shown in Donat's paper (in their third order simulation).
A smaller value, and a larger spreading, is apparent in Fig.~\ref{fig3},
where the most smearing case is tested, that is minmod limiter and
{\em global} Lax-Friedrichs solver with $\alpha=1$, corresponding
to $\lambda^\pm\to\pm 1$ (speed of light) in Eq.~(\ref{lambda}).
However, given the extreme simplicity of the scheme, even this case
should not be regarded as completely unusable.

\begin{figure}
\resizebox{\hsize}{!}{\includegraphics{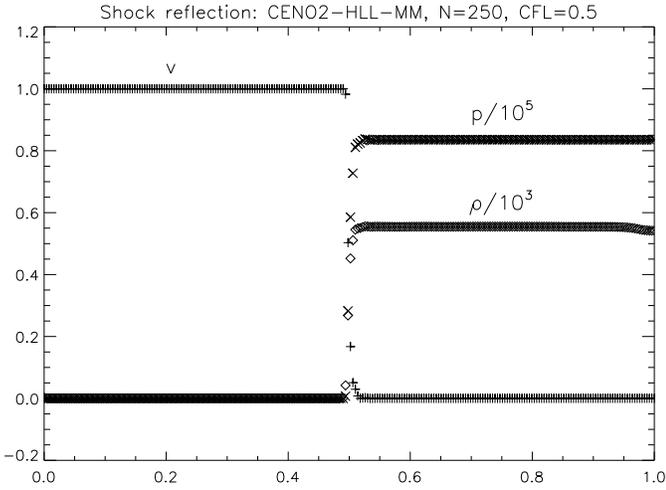}}
\caption{The relativistic shock reflection problem for time $t=0.75$.
The computed profiles of density (diamond), pressure (cross), and
velocity (plus) are shown against distance. Here the second order CENO2 
scheme is used with MM limiter and HLL solver. The computational grid 
is formed by $N=250$ equidistant cells and the Courant number is $c=0.5$.
Note that the error in the density due to the {\em wall heating} phenomenon
is around $2.3\%$, to be compared with the value of $2.5\%$ given by
Marquina's scheme in its {\em third order} implementation.}
\label{fig4}
\end{figure}

The next example considered is the notorious relativistic shock reflection
problem, where an ultrarelativistic cold wind hits a wall, 
a shock propagates backwards and a static region of relativistically hot
gas ($e\gg\rho\Rightarrow c_s^2\to \Gamma-1$) is left behind.
The numerical box is again $[0,1]$ (we use $N=250$) and the reflecting
wall is placed at $x=1$. The physical values employed are:
$$
\mbox{Shock reflection:}~~ (\rho,v,p)=(1,0.99999,0.01),
$$
that corresponds to a Lorentz factor as high as $\gamma\simeq 223$, 
about the highest allowed in single precision calculations.
This is a very severe test, for the high velocities involved and for the
so-called {\em wall heating} problem, visible as a dip in the density
profile near the reflecting wall. This is a classical numerical artifact,
due basically to the implicit numerical viscosity present in every scheme.
In Fig.~\ref{fig4} a simulation with second order accuracy (both in time
and space), HLL solver and MM limiter is shown for time $t=0.75$ ($c=0.5$).
This is the only case where the third order CENO3 has failed, giving
significant postshock oscillations, that can only be reduced by lowering
the Courant number or by enhancing numerical viscosity, though they never
completely disappear. However, the overheating error in the density
is here only $2.3\%$, in spite of the second order of accuracy and of
our simplified scheme, to be compared with the $2.5\%$ value of Donat's
paper in their third order implementation of the celebrated Marquina's
scheme.

\begin{figure}
\resizebox{\hsize}{!}{\includegraphics{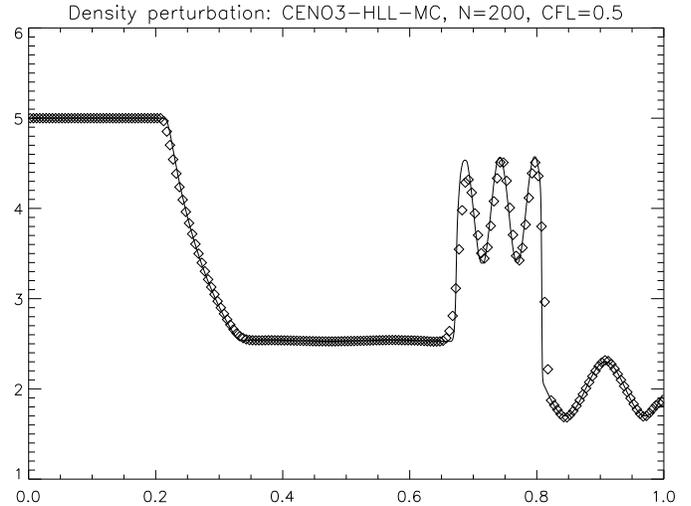}}
\caption{The Riemann problem with sinusoidal density perturbation for
time $t=0.35$. The diamonds are the results obtained with CENO3, MC
limiter and HLL solver, in a simulation with $N=200$ grid points and
Courant number is $c=0.5$. The solid line is the density profile that
comes out from a simulation with 2000 grid points. High-order ENO schemes
are particularly suited for problems involving shocks interacting with
smooth wave-like structures.}
\label{fig5}
\end{figure}

As a last 1-D application we present a test proposed in the Dolezal \&
Wong (1995) paper, which nicely shows the capacity of ENO-based schemes
of treating accurately both discontinuous and smooth features occurring
close together and at the same time. A shock tube is perturbed in its
right-hand state ($0.5<x\leq 1$) with a density sinusoidal profile:
$$
\mbox{Density perturbation:}
\left\{ \begin{array}{ll} 
(\rho,v,p)^L = & \!\! (5,0,50), \\
(\rho,v,p)^R = & \!\! (2+0.3\sin 50x,0,5).
\end{array}\right.
$$
The subsequent evolution in time shows the interaction of the blast
wave with the density wave (the values are not exactly the same as
in Dolezal's paper, where a nuclear equation of state is used), which
enters the expanding heated region and modulates its density plateau. 
In Fig.~\ref{fig5} the extreme
accuracy of our third order CENO3 scheme is shown, by comparing the
resulting density profiles at time $t=0.4$ in two runs, one with just 
$N=200$ points (diamonds), tested against one with $N=2000$ grid points
(solid line). Again, by comparing our results to those obtained by
the third order {\em characteristics based} ENO scheme in Dolezal's paper,
it is apparent that no significant differences arise, in spite of the
much less effort involved in our scheme.

\subsection{Multidimensional tests}

Multidimensional relativistic simulations are more difficult than
one-dimensional ones because the velocity components are spatially interpolated 
separately, possibly causing the condition $v^2<1$ to fail in ultrarelativistic
regimes due to numerical errors in the reconstruction.
For this reason in some cases we had to reduce to first order reconstruction,
namely when the Lorentz factor reaches $\gamma=10$. Note that this does
not mean in any sense that this threshold cannot be exceeded, but only
that the accuracy is lower in these regions, which usually are located
at shock fronts where the order would have been lowered anyway.
We have also tested reconstruction on four-velocity components, which
are not bounded, but in this way the problem is just shifted to the
primitive variables finding routine, thus with no improvement at all.

\begin{figure}
\resizebox{\hsize}{!}{\includegraphics{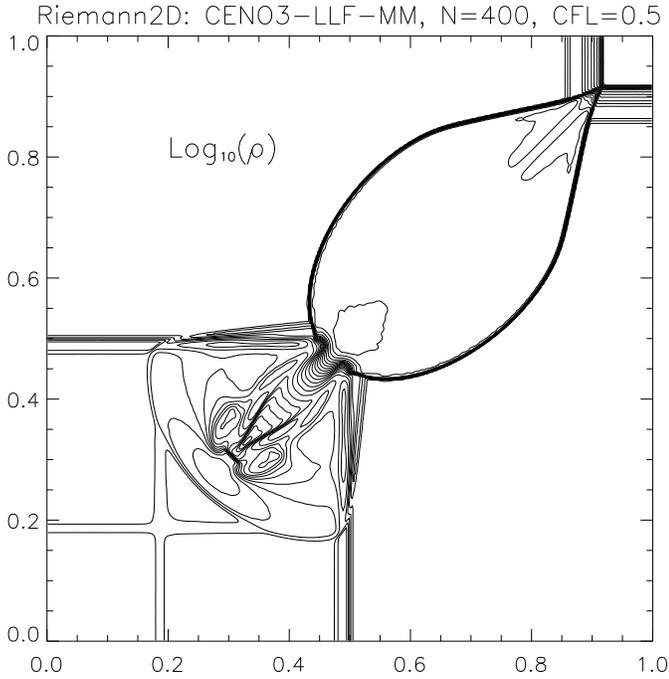}}
\caption{The relativistic 2-D Riemann problem for time $t=0.4$. 
Density contours (30) in logarithmic scale are shown, for a simulation
using CENO3, MM limiter and LLF solver, with $N_x=N_y=400$ grid points and
Courant number is $c=0.5$.}
\label{fig6}
\end{figure}

While it easy to test one-dimensional codes, since Riemann problems
can be solved exactly through iterative algorithms, it is not so in more
than one dimension, where it is rare to see really quantitative numerical
scheme validations. Here we will propose a 2-D Riemann problem, 2-D and
3-D explosions, compared with the correspondent 1-D cylindrical and
spherical solutions, and finally one simulation of a relativistic jet,
which is not a quantitative test but it is a true astrophysical application.

The two-dimensional counterpart of shock tubes is a square domain divided
in four quadrants of constant values at initial time and free evolution
for $t>0$. In Lax \& Liu (1998) all the possible different configurations
involving 1-D shocks, 1-D rarefactions waves and 2-D slip lines (contact
discontinuities) were studied in detail. In Fig.~\ref{fig6} we show the
output (contours of the density logarithm for time $t=0.4$) of a situation
similar to their configuration 12, where the four boundaries defines
two contact discontinuities and two 1-D shocks (symmetric with respect
to the main diagonal). Here a relativistic version of this test is proposed,
with the following initial settings:
$$
\mbox{Riemann 2-D:}
\left\{ \begin{array}{ll}
(\rho,v_x,v_y,p)^{NE} =  & (0.1,0,0,0.01), \\
(\rho,v_x,v_y,p)^{NW} =  & (0.1,0.99,0,1), \\
(\rho,v_x,v_y,p)^{SW} =  & (0.5,0,0,1),  \\
(\rho,v_x,v_y,p)^{SE} =  & (0.1,0,0.99,1).
\end{array}\right.
$$
Note that we have not taken exact 1-D shocks across the N and E
interfaces, and this may be recognized by observing
the evolved discontinuities in Fig.~\ref{fig6}, converging towards the
NE corner with their complete Riemann fans. In the rest of the domain 
the structure evolves with curved shock fronts and a complicated pattern
in the SW quadrant, reminiscent of an oblique jet with bow shock and
converging internal shock fronts. The lines in the SW direction with respect
to the bow shock are actually due to spurious waves created at the W and S
interfaces by the numerical diffusion term in the energy equation
(left and right states have a jump in kinetic energy), and cannot be
removed if not by using a Roe-type solver. Note that the most diffusive
case (LLF solver and MM limiter) was used in the simulation, since other
cases resulted more unstable at the same high resolution and with the
same Courant number $c=0.5$.

\begin{figure}
\resizebox{\hsize}{!}{\includegraphics{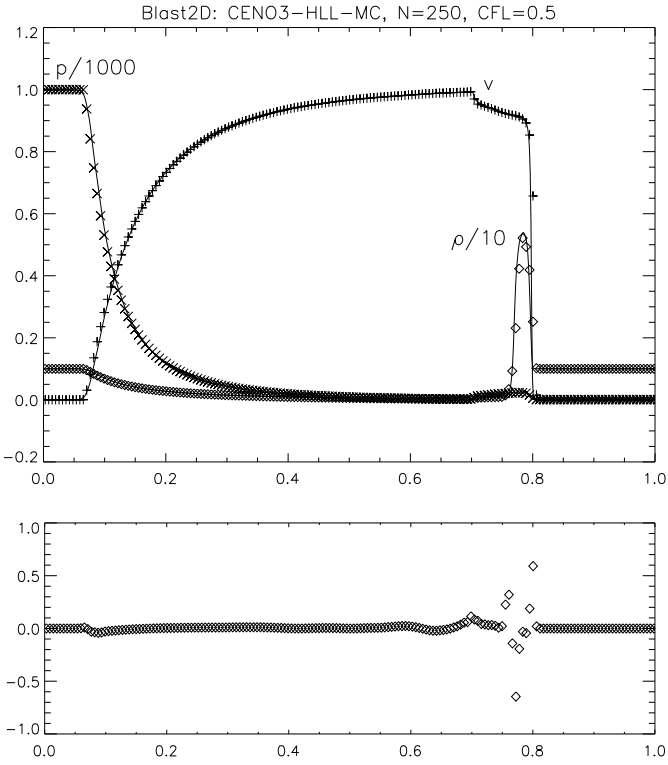}}
\caption{The relativistic 2-D blast wave problem for time $t=0.4$.
Here the simulation employed CENO3 reconstruction, MC limiter and HLL solver, 
with $N_x=N_y=N=250$ grid points. The domain is a Cartesian 2-D unit box 
$[0,1]\times [0,1]$, with initial enhanced pressure (a factor of $10^3$)
in a disk sector centered at the origin with radius $r_\mathrm{max}=0.4$
(reflecting boundary conditions are assumed for $x=0$ and $y=0$).
A radial cut (along the main diagonal) of the computed quantities
are compared with those obtained through a high-resolution ($N=800$)
1-D simulation in cylindrical coordinates (solid line), 
using the same parameters.
For a more quantitative comparison, in the bottom panel the density
relative error is plotted. Its rather large value near shocks is simply due
to the naturally reduced accuracy at discontinuities and to the very small
number of grid points (eight) present in the density peak.}
\label{fig7}
\end{figure}

\begin{figure}
\resizebox{\hsize}{!}{\includegraphics{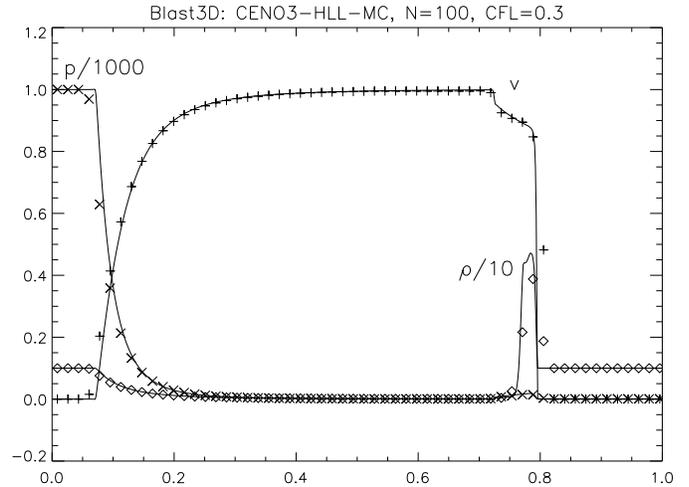}}
\caption{The relativistic 3-D blast wave problem for time $t=0.4$.
All the settings are the same as in Fig.~\ref{fig7}, except that here
$N=100$ along all directions and the Courant number is lower ($c=0.3$).
A radial cut (along the main diagonal) of the computed quantities
are compared with those obtained through a high-resolution ($N=800$)
1-D simulation in spherical coordinates (solid line), 
using the same parameters.}
\label{fig8}
\end{figure}

\begin{figure}
\resizebox{\hsize}{!}{\includegraphics{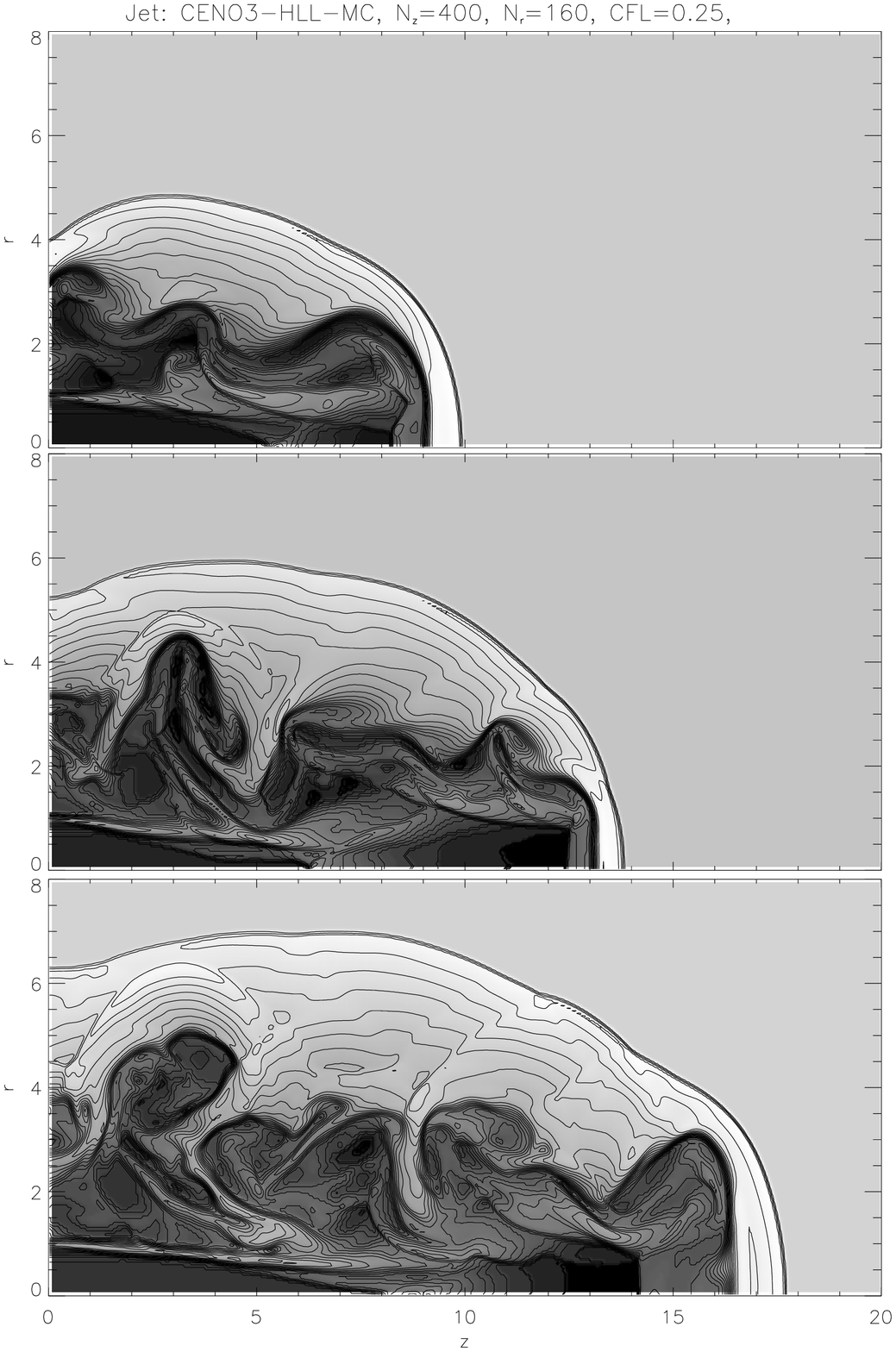}}
\caption{The relativistic jet for times $t=20,~30,~40$ in 2-D cylindrical 
geometry (logarithmic density contours and shades are shown).
The inflow speed is $v_z=0.99$ and the internal to external density ratio
is $\eta=1/100$, with an overall pressure equilibrium, leading to
a beam relativistic Mach number of $17.9$. The jet radius is assumed as
length unit, corresponding to $20$ computational cells ($N_z=400$ grid 
points in the axial direction and $N_r=160$ points in the radial direction). 
CENO3 reconstruction, HLL solver and MC limiter have been used in the 
simulation, with Courant number $c=0.25$.}
\label{fig9}
\end{figure}

Fig.~\ref{fig7} and Fig.~\ref{fig8} refer respectively to the 2-D 
cylindrically symmetric and 3-D spherically symmetric blast wave
explosion problem below:
$$
\mbox{Radial blast wave:}
\left\{ \begin{array}{lll} 
(\rho,v_r,p) = & (1,0,1000) & r\leq 0.4, \\
(\rho,v_r,p) = & (1,0,1) & r>0.4,
\end{array}\right.
$$
where in both cases the runs have been performed in a Cartesian unit box 
(with reflective right hand side boundary conditions along each direction)
and compared with the correspondent 1-D radial solution, obtained with
$N=800$ grid points in cylindrical or spherical coordinates.
The results are satisfactory, since the radial symmetry is well preserved
(here the results along the diagonal are shown) and high Lorentz factors
are reached without particular problems ($\gamma>25$ in the spherically
symmetric case). Here the scheme used is CENO3 with HLL solver and
MC limiter, with $c=0.5$ and $c=0.3$ in the 2-D and 3-D cases, respectively.

Finally, as a typical astrophysical test, we simulate the propagation
of an axisymmetric jet in 2-D cylindrical coordinates $(z,r)$. Note that
jet simulations are a very hard test for codes not based on characteristics
decomposition, because of usually stronger numerical viscosity at shear layers.
The domain is $0<r<8$ and $0<z<20$, with reflective boundary conditions 
on the axis $r=0$ and simple extrapolation at the other boundaries
(except at $z=0$ within the jet radius, where initial values are
kept constant).
At $t=0$ a relativistic jet with $v_z=0.99$ and density 100 times less
than the surroundings (but same pressure) is located at $r\leq 1$ and
$z\leq 1$ 
$$
\mbox{Jet:}
\left\{ \begin{array}{lll} 
(\rho,v_z,v_r,p) = & (0.1,0.99,0,0.01) & r\leq 1,z\leq 1, \\
(\rho,v_z,v_r,p) = & (10,0,0,0.01) & \mbox{outside};
\end{array}\right.
$$
density and velocity jumps are actually smoothed in order to reduce
spurious transverse waves that appear due to numerical viscosity
at the shear layer. The material is injected with a Lorentz factor
$\gamma\simeq 7.1$, corresponding to a relativistic Mach number 
(e.g. Duncan \& Hughes 1994) of ${\cal M}=\gamma v/\gamma_{c_s}c_s\simeq 17.9$.
The jet evolution is followed until $t=40$, as shown in Fig.~\ref{fig9},
where density contours and gray shades in logarithmic scale are presented.
The code settings are CENO3-HLL-MC with Courant coefficient $c=0.25$,
while the resolution employed is $400\times 160$, corresponding to 20
grid points per jet radius.
Note that the smearing of contact discontinuities, unavoidable in methods 
not based on characteristics decomposition, is actually small and vortices
due to Kelvin-Helmoltz instabilities are nicely defined, as well as the 
external bow shock, the internal Mach disk and other shocks reflected
off the axis. Moreover, notice the absence of the so-called {\em carbuncle}
problem, usually manifesting as an extended {\em nose} in the front
of the jet on the axis (e.g. Quirk 1994).

\section{Conclusions}

The central shock-capturing scheme of Londrillo \& Del Zanna (2000) is
extended to the relativistic case, here for ideal RHD flows  
and to RMHD in the next paper of the series. Compared to other schemes
proposed for relativistic astrophysical problems over the last 
decade in the literature, our method is extremely simple and efficient,
since no eigenvector decomposition and Riemann solvers are involved.
In spite of this, due to the high order accuracy achieved in smooth 
regions and to TVD limiting near discontinuities, provided by the CENO
reconstructions employed, our results compete with those obtained by
more sophisticated (but less computationally efficient) codes, 
even in 2-D and 3-D highly relativistic test problems.
Different geometries (cylindrical and spherical coordinates) and boundary 
conditions have also been tested.
All the numerical experiments have been run on a simple PC-Linux machine,
even the most demanding 3-D test with one million grid points, in order
to demonstrate the efficiency of our code.
We believe that our simple scheme can be successfully employed in many
relativistic simulations of astrophysical interest, either in the fluid
case and above all in the most demanding magnetic case, which will be the
subject of a forthcoming paper.

\begin{acknowledgements}
The authors thank P. Londrillo and R. Bandiera for valuable discussions.
This work has been partly supported by the Italian Ministry for University and
Research (MIUR) under grants Cofin2000--02--27 and Cofin2001--02--10.
\end{acknowledgements}

\appendix
\section{Convex ENO interpolation routines}

Given a function $v(x)$ with grid point values $\{v_i\}$, for $i=1,\ldots,N$,
let us see how non-oscillatory reconstructions ({\em Rec}) and second
derivatives ({\em Der}) are defined in our scheme. These procedures are
based on the original CENO reconstruction from Liu \& Osher (1998) and
they have been described also in LD (we repeat them here for completeness).

In the range $x_{i-1/2}\leq x_i\leq x_{i+1/2}$, we first consider the two
linear polynomials $L_i(x)$, based on the stencils $[x_{i-1},x_i]$ or
$[x_i,x_{i+1}]$, and we choose their convex combination which is closest
to $v_i$, that happens to coincide with the TVD limited reconstruction:
\be
\tilde{L}_i(x)=v_i+v_i^\prime \left(\frac{x-x_i}{\Delta x}\right).
\ee
Here the non-oscillatory undivided first derivative $v_i^\prime$ 
is given by the {\em minmod} function:
\be
v_i^\prime=\mathrm{mm}[\Delta_-v_i,\Delta_+v_i],
\ee
and $\Delta_\pm v_i=\pm (v_{i\pm 1}\pm 1)$.
Thus, to second order of accuracy the CENO technique simply reduces to
the usual TVD monotone slope limiting. Although not in the philosophy
of the original CENO approach, we have tested other limiters. The most
compressive ones, like {\em superbee} are too oscillatory in our
component-wise framework, while a good compromise between sharpening
and stability appears to be provided by the so-called 
{\em monotonized centered} (TVD for $1\leq\theta\leq 2$):
\be
v_i^\prime=\mathrm{mm}[\theta \Delta_-v_i,\Delta_0 v_i,\theta \Delta_+v_i],
\ee
where $\Delta_0\,v_i=(\Delta_+v_i+\Delta_-v_i)/2=(v_{i+1}-v_{i-1})/2$
and $\theta=2$ in order to bias towards central interpolation (since
twice the one-sided derivatives might be chosen near discontinuities, 
the Courant coefficient should be lowered correspondingly for overall 
stability). In both relations the function $\mathrm{mm}$ is defined as
\be
\mathrm{mm}(a_1,a_2,\ldots)=
\left\{ \begin{array}{ll}
\mathrm{min}_j\{a_j\}  & a_j\geq 0~~ \forall j, \\
\mathrm{max}_j\{a_j\}  & a_j\leq 0~~ \forall j, \\
0                      & \mathrm{otherwise}.
\end{array}\right.
\ee

To achieve an accuracy of third order in smooth regions, we need to 
calculate the three quadratic polynomials ($k=-1,0,+1$, $j=i+k$):
\be
Q_i^k(x)=v_j+\Delta_0\,v_j\left(\frac{x-x_j}{\Delta x}\right)+
\frac{1}{2}\Delta_0^2v_j\left(\frac{x-x_j}{\Delta x}\right)^2\!\!\!,
\ee
where $\Delta_0^2v_j=\Delta_+\Delta_-v_j=\Delta_-\Delta_+v_j=v_{j+1}
-2v_j+v_{j-1}$, together with the corresponding weighted differences
\be
d^k_i=\alpha^k(Q^k_i-\tilde{L}_i).
\ee
Then, we take again their convex
combination which is closest to $\tilde{L}_i$ at a given point $x$
(this procedure is the generalization of {\em minmod} limiting
to higher orders) and finally, in smooth regions where all $d^k_i$ have
the same sign, we take
\be
\tilde{Q}_i=Q_i^{k_0},~~|d_i^{k_0}|=\mathrm{min}_k(|d_i^k|),
\ee
otherwise we exit from the interpolation routine with 
$\tilde{Q}_i=\tilde{L}_i$.
As for the coefficient $\theta$ in the limiter of the lower order, 
the weights $\alpha^k$ are chosen in order
to bias towards central interpolation, and following the original
recipe for component-wise CENO schemes we take $\alpha^{\pm 1}=1$
and $\alpha^0=0.7$.
In our {\em Rec} routine, the left and right states are calculated at
the same time and the output is finally
\be
v_{i+1/2}^L=\tilde{Q}_i(x_{i+1/2}),~~v_{i-1/2}^R=\tilde{Q}_i(x_{i-1/2}).
\ee

The other CENO routine used in our scheme is {\em Der}, which allows
one to calculate non-oscillatory second derivatives 
${\bar{v}}_i^{\prime\prime}$, needed to transform
cell averages into point values, calculated at the same point:
\be
\{\bar{v}_i\}\longrightarrow\{v_i\},~~v_i=\bar{v}_i-
\frac{1}{24}{\bar{v}}_i^{\prime\prime}.
\ee
In this case, the above CENO selection criterion applies with null
linear polynomials, $\tilde{L}_i=0$, and quadratic ones given by
($k=-1,0,+1$, $j=i+k$):
\be
Q_i^k=\Delta_0^2\,\bar{v}_j\equiv\bar{v}_{j+1}-2\bar{v}_j+\bar{v}_{j-1},
\ee
then finally ${\bar{v}}_i^{\prime\prime}=\tilde{Q}_i$ (which is zero,
so $\bar{v}_i\equiv v_i$ as for second order approximation, in the
case different signs of the $Q_i^k$ terms).


\begin{thebibliography}{}

\bibitem{} Aloy, M.A., Iba\~nez, J.M., Marti, J.M., M\"uller, E., 1999,
           ApJS, 122, 151

\bibitem{} Anile, M., 1989, Relativistic Fluids and Magneto-Fluids
           (Cambridge University Press, Cambridge)

\bibitem{} Balsara, D.S., 1994, J. Comput. Phys., 114, 287

\bibitem{} Balsara, D.S., 2001, ApJS, 132, 83

\bibitem{} Begelman, M.C., Blandford, R.D., Rees, M.J., 1984, Rev. Mod.
           Phys., 56, 255

\bibitem{} Bucciantini, N., Bandiera, R., 2001, A\&A, 375, 1032

\bibitem{} Dolezal, A., Wong, S.S.M., 1995, J. Comput. Phys., 120, 266

\bibitem{} Donat, R., Font, J.A., Iba\~nez, J.M., Marquina, A., 1998,
           J. Comput. Phys., 146, 58

\bibitem{} Duncan, G.C., Hughes, P.A., 1994, ApJ, 436, L119

\bibitem{} Eulderink, F., Mellema, G., 1994, A\&A, 284, 654

\bibitem{} Falle, S.A.E.G., Komissarov, S.S., 1996, MNRAS, 278, 586

\bibitem{} Ferrari, A., 1998, ARA\&A, 36, 539

\bibitem{} Font, J.A., Iba\~nez, J.M., Marquina, A., Marti, J.M., 1994,
           A\&A, 282, 304

\bibitem{} Harten, A., Engquist, B., Osher., S., Chakravarthy, S., 1987,
           J. Comput. Phys., 71, 231

\bibitem{} Jiang, G., Shu, C.-W., 1996, J. Comput. Phys., 126, 202

\bibitem{} Kennel, C.F., Coroniti, F.V., 1984, ApJ, 283, 694

\bibitem{} Kobayashi, S., Piran, T., Sari, R., 1999, ApJ, 513, 669

\bibitem{} Komissarov, S.S., 1999, MNRAS, 303, 343

\bibitem{} Kurganov, A., Noelle, S., Petrova, G., 2001,
           SIAM J. Sci. Comput., 23, 707

\bibitem{} Landau, L.D., Lifshitz, E.M., 1959, Fluid Mechanics
           (Pergamon Press)

\bibitem{} Lax, P.D., Liu, X.-D., 1998, SIAM J. Sci. Comput., 19, 319

\bibitem{} Liu, X.-D., Osher, S., 1998, J. Comput. Phys., 142, 304

\bibitem{} Londrillo, P., Del Zanna, L., 2000, ApJ, 530, 508 (LD)

\bibitem{} Marquina, A., Marti, J.M., Iba\~nez, J.M., Miralles, J.A.,
           Donat, R., 1992, A\&A, 258, 566

\bibitem{} Meszaros, P., Rees, M.J., 1992, MNRAS, 258, 41

\bibitem{} Michel, F.C., Li, H., 1999, Phys. Rep., 318, 227

\bibitem{} Mirabel, I.F., Rodriguez, L.F., 1999, ARA\&A, 37, 409

\bibitem{} Nessyahu, H., Tadmor, E., 1990, J. Comput. Phys., 87, 408

\bibitem{} Piran, T., 1999, Phys. Rep., 314, 575

\bibitem{} Quirk, J., 1994, Int. J. Numer. Methods Fluids, 18, 555

\bibitem{} Rees, M.J., 1967, MNRAS, 137, 429

\bibitem{} Rees, M.J., Gunn, J.E., 1974, MNRAS, 167, 1

\bibitem{} Schneider, V., Katscher, U., Rischke, D.H., Waldhauser, B.,
           Maruhn, J.A., Munz, C.-D., 1993, J. Comput. Phys., 105, 92

\bibitem{} Shu, C.-W., 1997, ICASE Rep. 97-65, NASA Langley Research Center, 
           VA

\bibitem{} Shu, C.-W., Osher, S., 1988, J. Comput. Phys., 77, 439

\bibitem{} Shu, C.-W., Osher, S., 1989, J. Comput. Phys., 83, 32

\end{thebibliography}
\end{document}